\documentclass[a4paper]{spie}  %>>> use this instead for A4 paper
\usepackage[]{graphicx}

\usepackage{psfig}
\usepackage{colordvi}
\title{\Large\bf Improving XMM--Newton EPIC pn data at low energies:\\
method and application to the Vela SNR}

\author{Konrad Dennerl\supit{1}, Bernd Aschenbach\supit{1},
Ulrich G. Briel\supit{1}, Hermann Brunner\supit{1},
Vadim Burwitz\supit{1}, Jakob Englhauser\supit{1,2},
Michael J. Freyberg\supit{1,3}, Frank Haberl\supit{1},
Gisela Hartner\supit{1,3}, Anatoli~F.~Iyudin\supit{1},
Eckhard Kendziorra\supit{4}, Norbert Meidinger\supit{1,2},
Elmar Pfeffermann\supit{1},
Wolfgang Pietsch\supit{1}, Lothar~Str{\"u}der\supit{1,2},
Vyacheslav E. Zavlin\supit{1,5}
\skiplinehalf\normalsize
\supit{1} Max--Planck--Institut f\"ur extraterrestrische Physik,
         Giessenbachstra{\ss}e, 85\,748 Garching, Germany
\skiplinehalf
\supit{2} MPI--Halbleiterlabor, Otto--Hahn--Ring 6, 81\,739 M\"unchen, Germany
\skiplinehalf
\supit{3} MPI--R\"ontgentestanlage PANTER, Gautinger Stra{\ss}e 45,
82\,061 Neuried, Germany
\skiplinehalf
\supit{4} Institut f\"ur Astronomie und Astrophysik, Sand 1,
72\,076 T\"ubingen, Germany
\skiplinehalf
\supit{5} Observatoire Astronomique, 11 rue de l'Universite,
67\,000 Strasbourg, France
}

\authorinfo{Further author information: (Send correspondence to K.D.)~
K.D.: E-mail: kod@mpe.mpg.de}
\begin{document} 
  \maketitle 

%%%%%%%%%%%%%%%%%%%%%%%%%%%%%%%%%%%%%%%%%%%%%%%%%%%%%%%%%%%%% 
\begin{abstract}
High quantum efficiency over a broad spectral range is one of the main
properties of the EPIC pn camera on--board XMM-Newton. The quantum efficiency
rises from $\sim75\,\%$ at 0.2~keV to $\sim100\,\%$ at 1~keV, stays close to
100\,\% until 8~keV, and is still $\sim90\,\%$ at 10~keV
\cite{99xmm015}. The EPIC pn camera is attached to an X--ray
telescope which has the highest collecting area currently available, in
particular at low energies (more than 1400~cm$^2$ between 0.1 and
2.0~keV)\cite{01aap296}. Thus, this instrument is very sensitive to
the low--energy X--ray emission. However, X--ray data at energies below
$\sim0.2\mbox{ keV}$ are considerably affected by detector effects, which
become more and more important towards the lowest transmitted energies. In
addition to that, pixels which have received incorrect offsets during the
calculation of the offset map at the beginning of each observation, show up as
bright patches in low--energy images. Here we describe a method which is not
only capable of suppressing the contaminations found at low energies, but
which also improves the data quality throughout the whole EPIC pn spectral
range. This method is then applied to data from the Vela supernova remnant.
\end{abstract}

%>>>> Include a list of keywords after the abstract 


%%%%%%%%%%%%%%%%%%%%%%%%%%%%%%%%%%%%%%%%%%%%%%%%%%%%%%%%%%%%%
\section{INTRODUCTION}
\label{sect:intro}  % \label{} allows reference to this section

\medskip
The EPIC pn camera on--board XMM--Newton is currently the most sensitive
instrument in space for detecting soft X--rays. In order to fully utilize this
unique capability, we have developed methods which improve the data quality in
particular at low energies. These methods are generally available in
SASS~v6.0, the most recent version of the XMM--Newton Science Analysis System,
in the task {\tt epreject}.

In Section~\ref{sect:offset} we describe a method which removes bright patches
that become visible in low--energy images. This is achieved by correcting the
energy of events which are detected in the corresponding pixels. Thus,
this method does not only improve the cosmetic quality of low--energy images,
but does also improve the spectral quality thoughout the whole bandpass of
EPIC pn. Although the most straightforward application of this method requires
the presence of the offset map for the corresponding exposure, we have extended
this method so that it can also be applied if only the event file is available.

Another improvement, described in Section~\ref{sect:detnoise}, concerns the
suppression of detector noise, which becomes significant at energies below
$\sim200\mbox{ eV}$. This is done by computing the probability that a specific
event may be a noise event and by assigning, on a statistical basis,
appropriate flags to individual events. Subsequent removal of events which
were flagged as due to detector noise leads to a conspicuous improvement of
the data quality below $\sim300\mbox{ eV}$ and makes the background
subtraction in this energy range more reliable. Additionally, as a
by--product, the event files become considerably smaller and easier to handle.

In Section~\ref{sect:vela}, we demonstrate these methods on the example of an
XMM EPIC pn observation of an area within the Vela supernova remnant. In
particular we show how the noise suppression method reveals the presence of
soft diffuse emission at energies below $\sim200\mbox{ eV}$ which would
otherwise be hidden in the detector noise.

\medskip
\section{OFFSET SHIFTS}
\label{sect:offset}

\subsection{The Problem}
\label{sect:problem}

\medskip
The usual mode of operating the EPIC pn camera consists of taking an offset
map immediately before the beginning of an exposure. Ideally, this map
contains for each pixel the energy offset (expressed in analog--to--digital
units, adu). During the subsequent exposure, these offsets are subtracted
on--board from the measured signals, and only events where the difference signal
exceeds a lower threshold (usually 20~adu) are transmitted to Earth.

\smallskip
The offset map is calculated in the following way: the charge in each pixel is
read out one hundred times, the three smallest and the three highest adu
values are ignored, and the sum of the remaining values, divided by 94 and
truncated to the nearest integer, is computed as the offset of that pixel.
This procedure is applied subsequently to regions of $64\times4\mbox{ pixels}$
along each CCD. The reason for excluding the extreme values is to minimize the
influence of energetic particles on the offset calculation. The highest adu
values are ignored because the charge deposited by {\em minimum}\/ ionizing
particles is usually close to the upper end of the dynamic adu range. However,
the charge deposited by a {\em highly}\/ ionizing particle may show up as a
very small charge, because the large amount of charge generated by such a
particle takes a long time to be cleared from the anode. Depending on the
number of signal electrons, the clearing goes on during the readout of the
next pixels, and the measured adu value is thus slighly reduced. Therefore,
also the smallest adu values are ignored.

%-------------------------------------------------------------------------------
\begin{figure}
 \hbox to \hsize{\hfil{\psfig{file=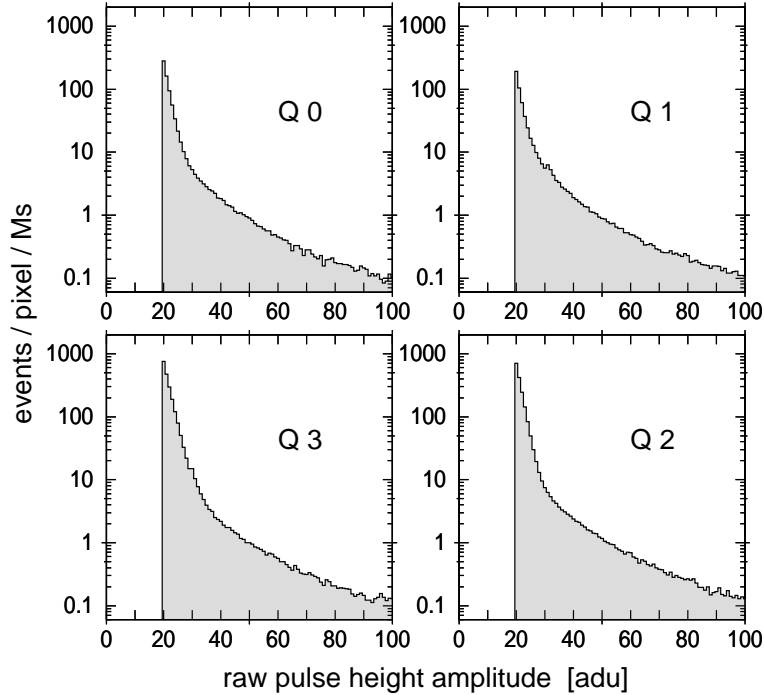,clip=,width=0.6\hsize,%
                    bbllx=84pt,bblly=221pt,bburx=546pt,bbury=652pt}}\hfil
\raisebox{20mm}{\vbox{\hsize=0.3\hsize
\caption{Number of events per pixel and Ms ($=10^6\mbox{ s}$) as a function of
raw pulse height amplitude [adu] for a $23.1\cdot10^3\mbox{ s}$ exposure
(obtained during revolution \#462) in full frame mode with the filter wheel
closed. These spectra are shown individually for all EPIC pn quadrants
(arranged according to their position on the detector). Note the different
intensity of detector noise in the four quadrants.
\label{offmp1a}
}}}\hfil}
\end{figure}
%-------------------------------------------------------------------------------

\begin{figure}
\psfig{file=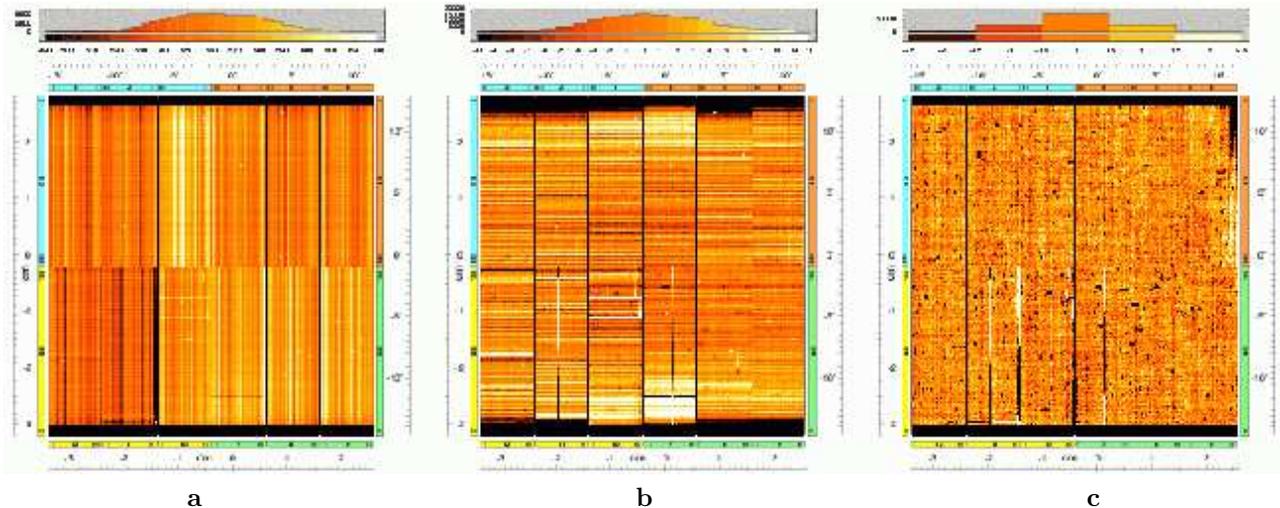,width=0.99\hsize,clip=}
\hbox to \hsize{\bf \hfil a \hspace*{55mm} b \hspace*{55mm} c \hfil}
\caption{Computing the ``residual offset map'': {\bf a)} Original offset map for
the observation 0526\_0147511101. The vertical stripes are mainly caused by
differences in the gain of the individual columns. {\bf b)} Same as (a), but
after subtracting from each column its median value. The horizontal stripes
are mainly caused by the common mode effect
(see Sect.\,\ref{sect:with_offset_map}).\
{\bf c)}~Same as (b), but after subtracting also from each row its median
value. The remaining patterns in this ``residual offset map'' are a
superposition of persistent features and temporary patches caused by energetic
particles.
\label{offset_prep_526}
}
\end{figure}

\smallskip
There are cases where so much charge is deposited by highly ionizing particles
that it takes several frames until the original state is restored. It may also
happen that several minimum ionizing particles are hitting the same pixels
during the one hundred readouts. At such occasions the technique of rejecting
the three smallest and the three highest adu values before computing the mean
offset is not sufficient, and the offset is not computed correctly. As the
erroneous adu values are predominantly too low, the computed mean offset
becomes too small. This effect introduces in the affected pixels deviations
from the correct offsets which range typically from $-5\mbox{ adu}$ to
$+3\mbox{ adu}$, corresponding to energy shifts from $-25\mbox{ eV}$ to
$+15\mbox{ eV}$. As a consequence, the energies of all events in these pixels
appear to be shifted by the same amount. The number of affected pixels scales
with the intensity of the particle background during the calculation of the
offset map and depends also on the readout mode. From our analysis of full
frame mode (FF) data, we find that 54\% of all pixels contain correct offsets,
while for 26\% the offset is by 5~eV too high and for 12\% it is by 5~eV too
small. Deviations of this amount ($\pm1\mbox{ adu}$) are not necessarily all
caused by energetic particles, as they may also reflect statistical and
numerical fluctuations in the computation of the offset map. Larger
deviations, however, can clearly be attributed to energetic particles, as can
be seen by their characteristic pattern (e.g.\ Fig.\,\ref{offset_prep_526}c).
For 5\% the offset is by 10~eV too high and for 2\% it is by 10~eV too small.
Offsets of 15~eV and more were found for 1\% of the pixels. For $\sim70\%$ of
them, the offset was too small.

%-------------------------------------------------------------------------------
%-------------------------------------------------------------------------------

%-------------------------------------------------------------------------------
\begin{figure}
\psfig{file=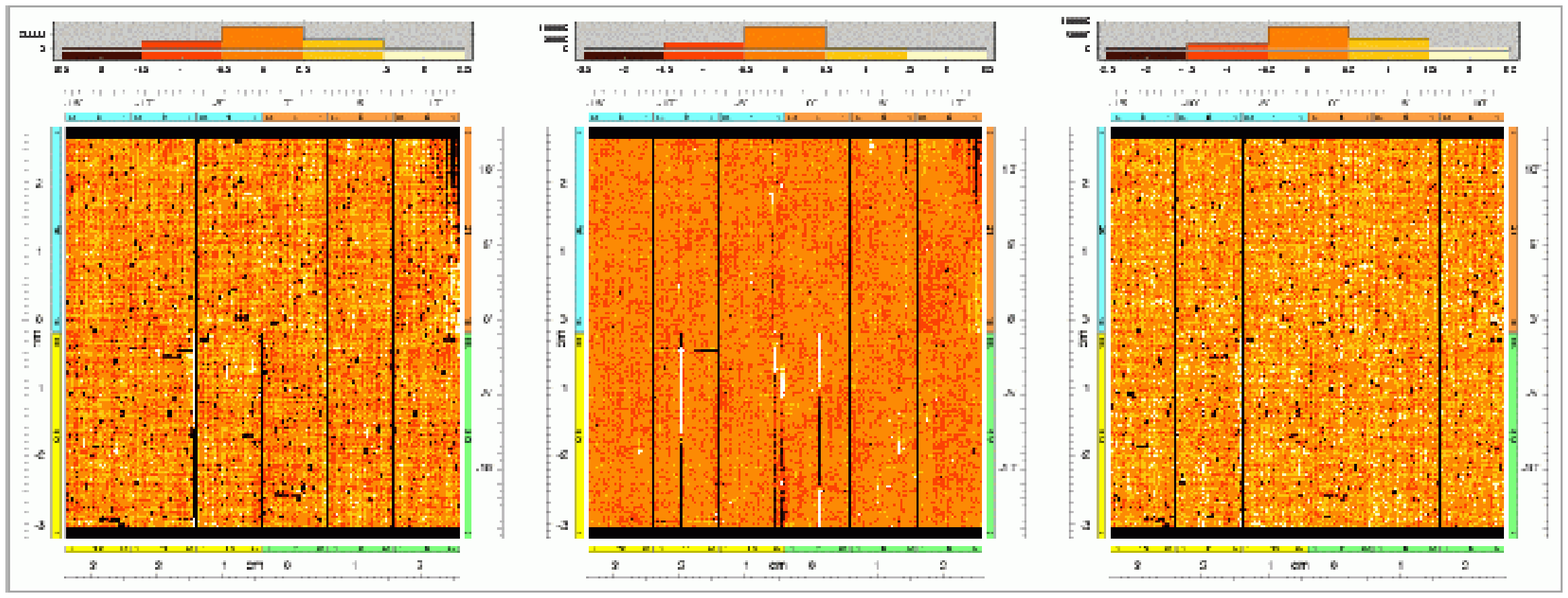,width=0.99\hsize,clip=}
\hbox to \hsize{\bf \hfil a \hspace*{55mm} b \hspace*{55mm} c \hfil}
\caption{Separating the persistent and temporary features in the
residual offset map: {\bf a)}
Same as Fig.\,\ref{offset_prep_526}\,c, but for a different observation
(523\_0147510801); note the different distribution of most patches.
{\bf b)} ``Residual offset reference map'': this map shows for each pixel the
median of seven residual offset maps (from revs 522, 523, 525, 526, 527, 528,
and 544) and thus contains the persistent features in the residual offset maps.
{\bf c)} Difference (a) $-$ (b), containing the temporary features for this
specific observation.
\label{offset_comp}
}
\vspace*{10mm}
\psfig{file=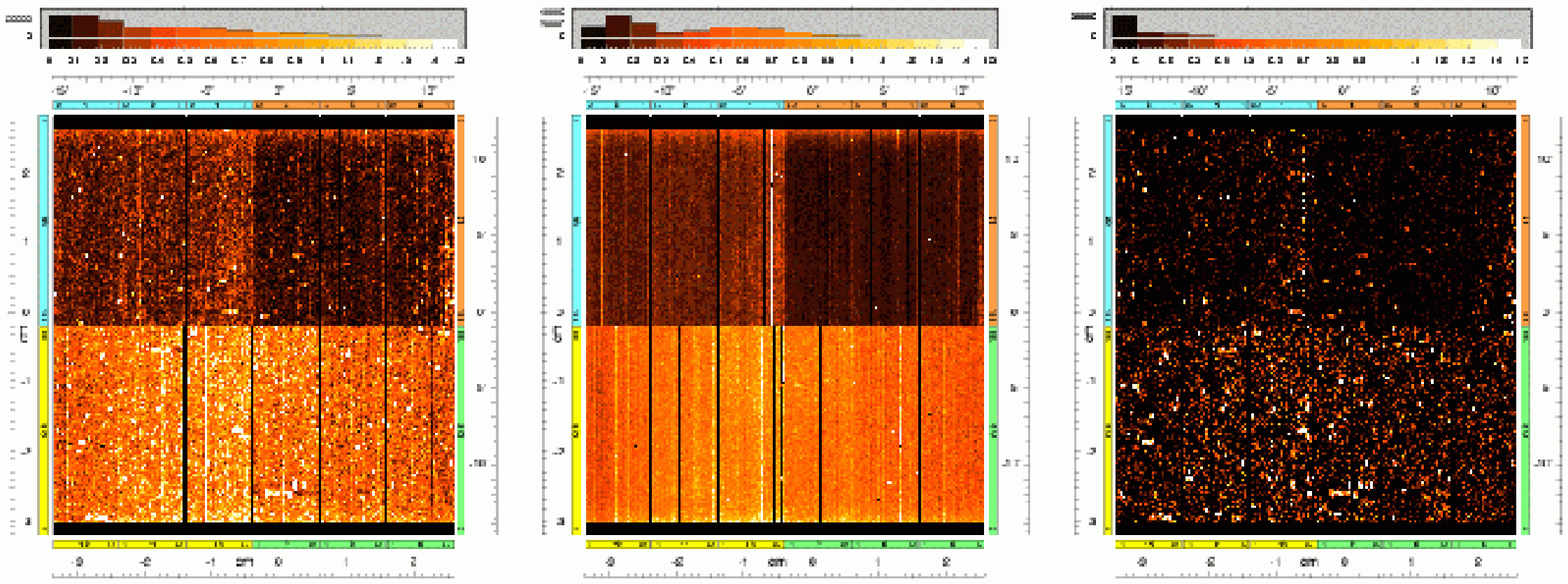,width=0.99\hsize}
\hbox to \hsize{\bf \hfil a \hspace*{55mm} b \hspace*{55mm} c \hfil}
\caption{Separating the persistent and temporary features in the 20~adu
image: {\bf a)} Distribution of all events with 20~adu accumulated in the 76~ks
observation 523\_0147510801, normalized to 1~ks. {\bf b)} ``20~adu reference
map'': this map shows for each pixel the median of ten residual offset maps
(from revs 344, 349, 522, 523, 525, 526, 527, 528, 544, and 548), each
normalized to 1~ks exposure, and thus contains the persistent features in the
20~adu maps. {\bf c)} Difference (a) $-$ (b), containing the temporary
features for this specific observation. Note the similarity with
Fig.\,\ref{offset_comp}\,c.
\label{ima20adu_comp}
}
\end{figure}
%-------------------------------------------------------------------------------

\smallskip
If the offset subtracted on--board was too small, then the adu values which are
assigned to events in such pixels became too high. Thus, events which have adu
values below the lower threshold and which would normally have been rejected,
may show up in the data set. As most of such events are due to detector noise,
which is steeply increasing towards lower energies (Fig.\,\ref{offmp1a}), any
reduction of the lower energy threshold leads to a considerable increase in
the number of events. Thus, pixels where the offset is too small show up as
bright patches in images accumulated in the lowest transmitted adu range. Due
to the specific pixel block sampling applied for the offset map calculation
(see above), these patches appear as rectangular areas, which are
characterized by a brightening in four consecutive pixels along readout
direction. Depending on the width of the trail caused by the energetic
particles and its orientation with respect to the CCD, the patches may also
extend over several consecutive pixels perpendicularly to the readout
direction. Pixels with too high offsets are less obvious in low--energy
images, as they are only somewhat darker than their environment.

\smallskip
While the occurrence of bright patches in EPIC pn images which are accumulated
at low energies (e.g.\,Fig.\,\ref{ima20adu_comp}) is immediately obvious,
there is also another consequence: a shift in the energy scale over the whole
spectral bandpass. This shift degrades the energy resolution for extended
objects. For point sources, the X--ray spectrum may be shifted by some 10~eV,
in most cases towards higher energies, if the position of the source happens
to coincide with one of these patches. Here we describe a method for restoring
the correct energy scale and for removing the bright patches in the soft
energy images. Although this paper concentrates on the low energy properties
of EPIC pn data, we note that the correction of the offset shifts improves the
energy calibration throughout the full spectral bandpass of the EPIC pn
detector.

\medskip
\subsection{Correcting the energy scale}

\smallskip
The correction is more straightforward if the offset map is available. If
this is not the case, then the offsets have to be reconstructed from the
spatial and spectral distribution of the transmitted events. These two cases
are described in the next sections.

\medskip
\subsubsection{If the offset map is available}
\label{sect:with_offset_map}

\smallskip
The offset map contains for each pixel the adu value which was subtracted from
all events in the particular pixel before transmitting this information to
ground. As it contains both correct and incorrect offsets, the main goal is to
distinguish between both cases. Such a distinction is possible under the
assumption that the generic offsets of all pixels (not affected by electronic
effects during readout) are very similar and that the areas where incorrect
offsets were computed occur in isolated patches.

\smallskip
The most obvious property of the offset map is a pattern of vertical stripes
(Fig.\,\ref{offset_prep_526}a). This is caused by the fact that each of the
768 readout columns has its own amplifier, and the amplification varies by
$\pm4.1\%$ (rms) from column to column. In order to spot the patches where
wrong offsets have been applied, the column to column variation has to be
removed. This can be achieved by subtracting the median value from each
column. The resulting map (Fig.\,\ref{offset_prep_526}b) is dominated by row
to row variations, which are mainly caused by the ``common mode" effect: a
temporal change of the offsets of all the pixels in a CCD row, occurring in a
predominantly irregular way. This effect is present during the calculation of
the offset map as well as during the subsequent exposure. During the exposure,
it is suppressed on--board by subtracting first the adu offsets contained in
the offset map, and then the median of all the adu values in a CCD row from
these values, individually for each readout frame. Concerning the further
processing of the offset map, this implies that the median value of each row
should also be subtracted from the corresponding row, in order to remove the
row to row variation. This leads to a ``residual offset map"
(Fig.\,\ref{offset_prep_526}c), which is close to zero except for the pixels
where specific offsets were applied.

%-------------------------------------------------------------------------------
\begin{figure}
\hbox to \hsize{%
  \psfig{file=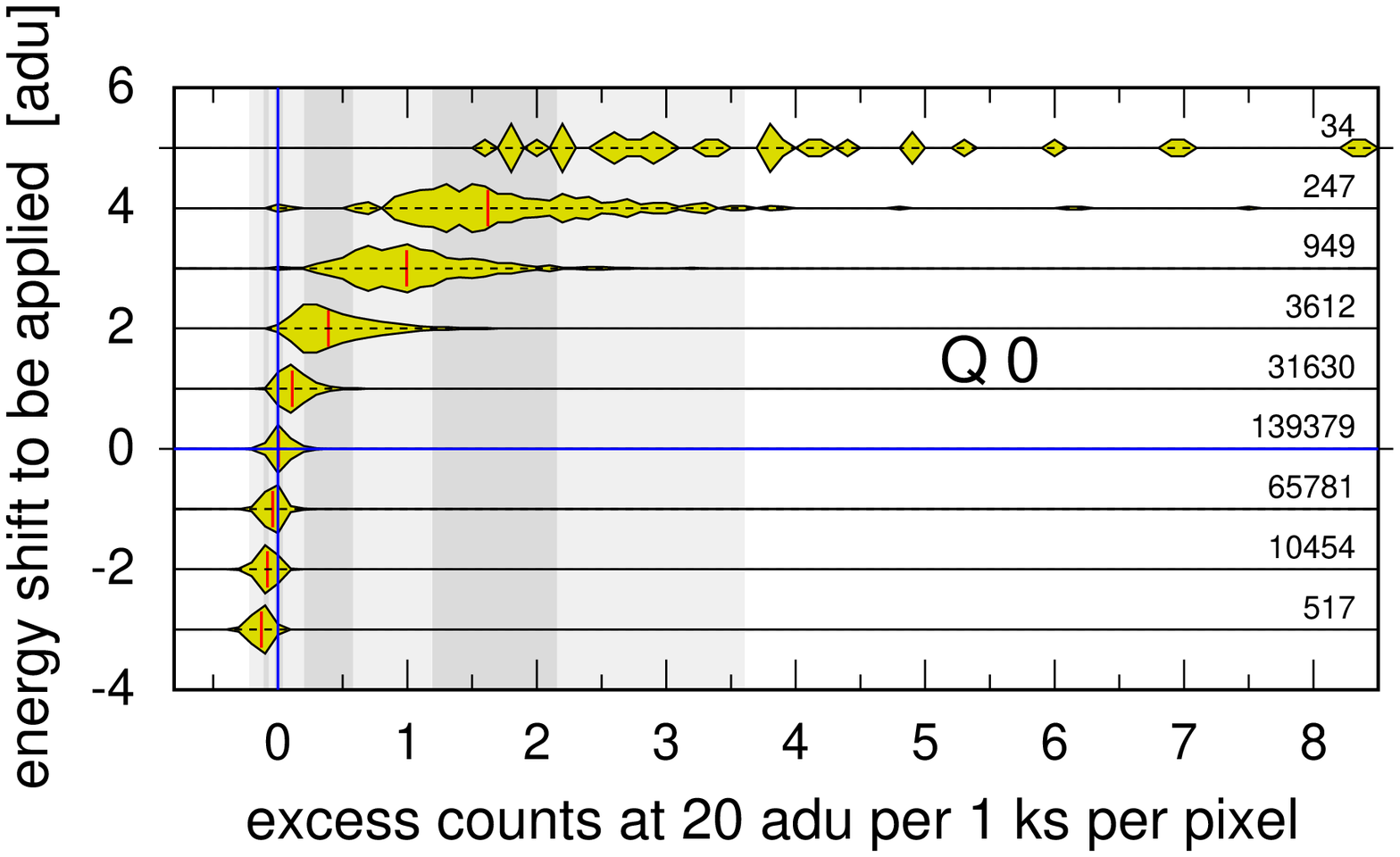,clip=,width=0.49\hsize,%
         bbllx=25pt,bblly=10pt,bburx=558pt,bbury=342pt}
  \hfil
  \psfig{file=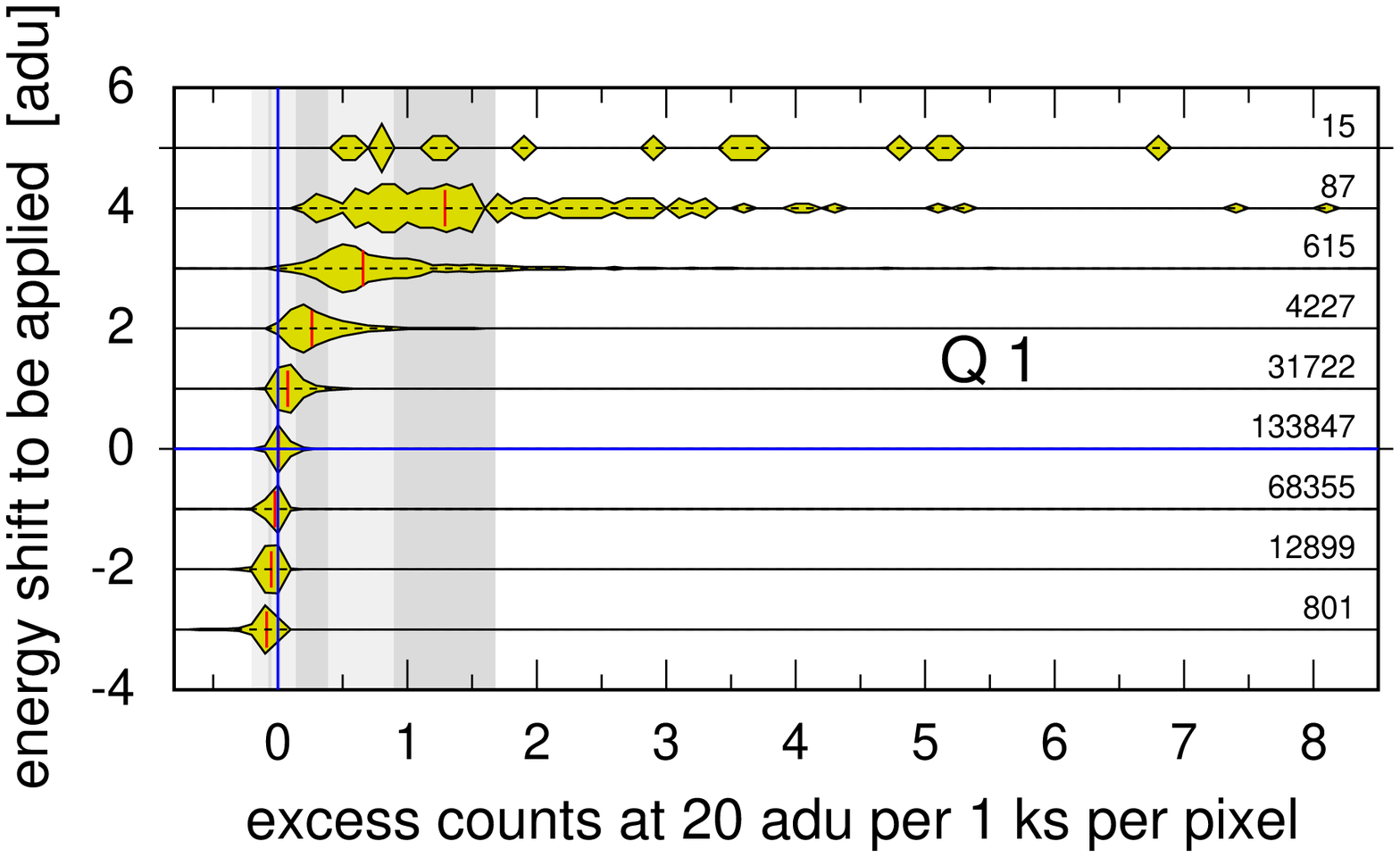,clip=,width=0.49\hsize,%
         bbllx=25pt,bblly=10pt,bburx=558pt,bbury=342pt}}
\hbox to \hsize{%
  \psfig{file=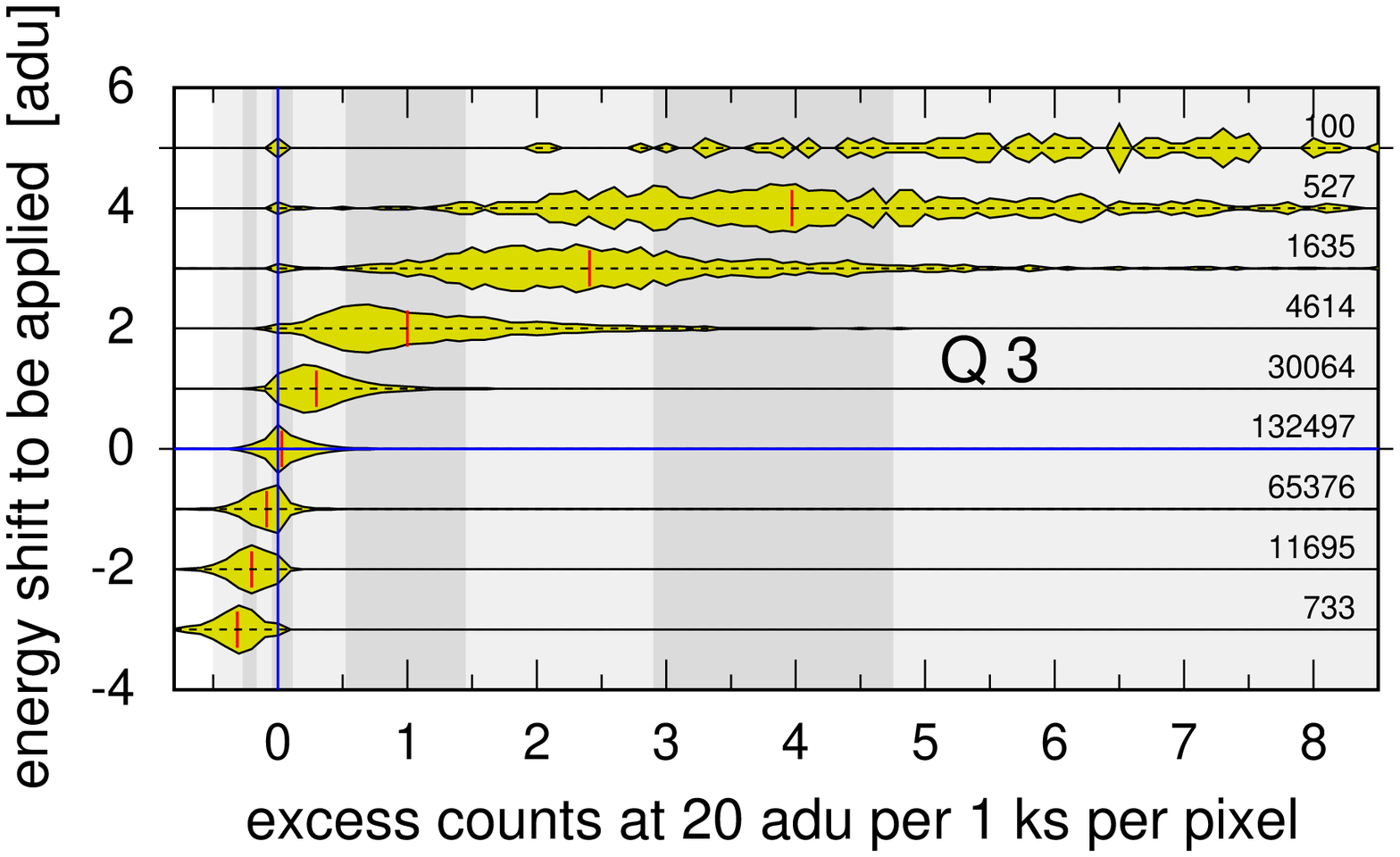,clip=,width=0.49\hsize,%
         bbllx=25pt,bblly=10pt,bburx=558pt,bbury=342pt}
  \hfil
  \psfig{file=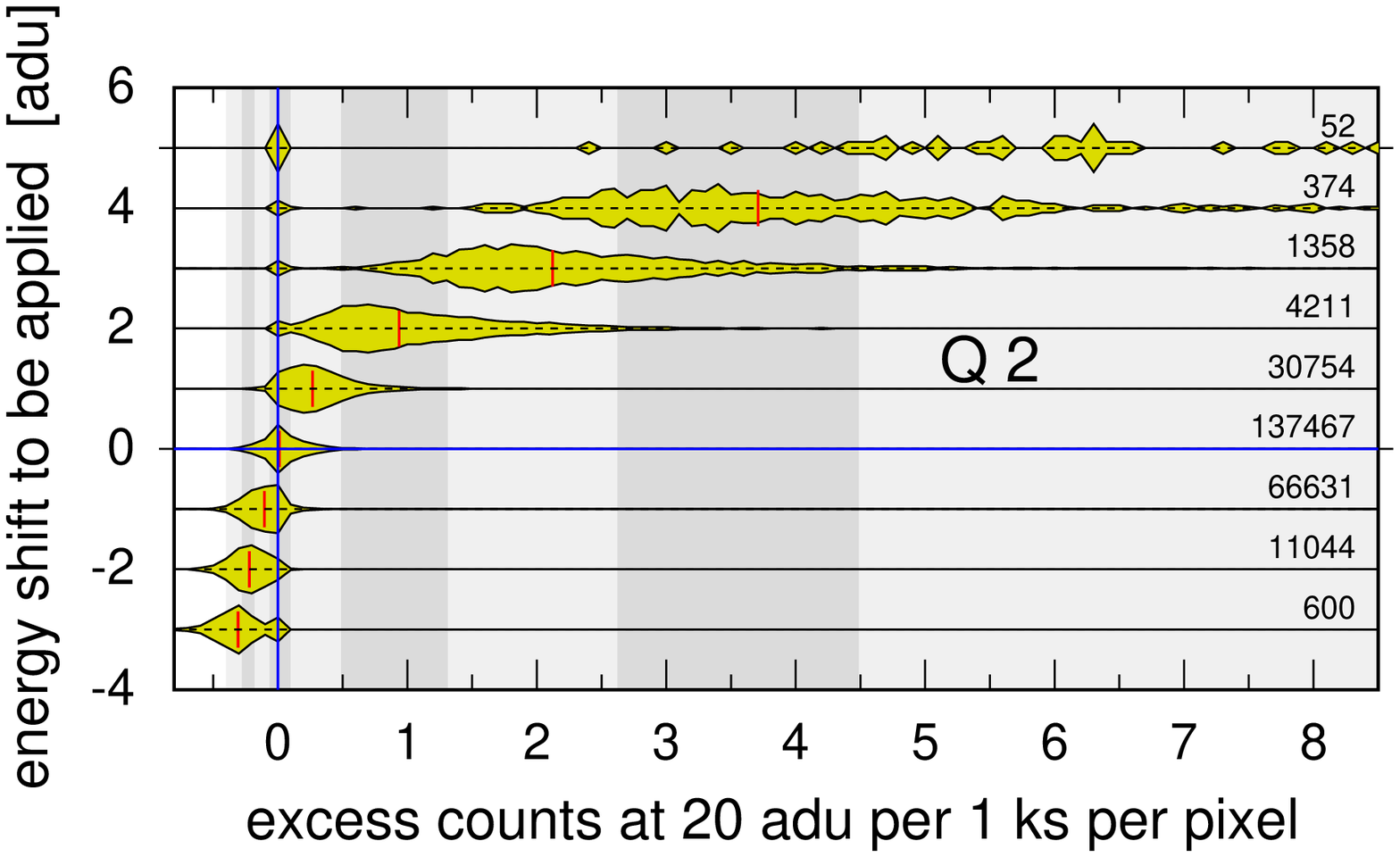,clip=,width=0.49\hsize,%
         bbllx=25pt,bblly=10pt,bburx=558pt,bbury=342pt}}
\vspace*{4mm}              
\caption{Pixel by pixel correlation between the excess counts at 20~adu,
normalized to 1~ks, and the values in the cleaned offset map,
for all quadrants (arranged according to their position on the detector).
While the energy shifts to be applied can only occur as integer multiples of
1~adu ($\sim5\mbox{ eV}$), the excess counts exhibit a much smoother
distribution due to their normalization to 1~ks. Their distribution is
illustrated by the thickness of ``histogram tubes'', which were all expanded to
have the same maximum thickness. The number of pixels which they contain are
given at right. This correlation was derived by combining the observations of
the Lockman hole in XMM revolutions 522, 523, 525, 526, 527, 528, and 544
($\sim636\mbox{ ks}$ total exposure time), all taken in fullframe mode.
For Q\,3, four noisy columns were
excluded. A vertical red line indicates the median value of the distributions.
Shaded areas show the ranges of excess counts for which a certain energy shift
should be applied. Note how the brightness of pixels at 20~adu responds
differently in the four quadrants to energy shifts. No such variation was
found within the CCDs of the same quadrant.
\label{offmp9_3_}
}
\end{figure}
%-------------------------------------------------------------------------------

%-------------------------------------------------------------------------------
\begin{figure}
\hbox to \hsize{\hfil
\psfig{file=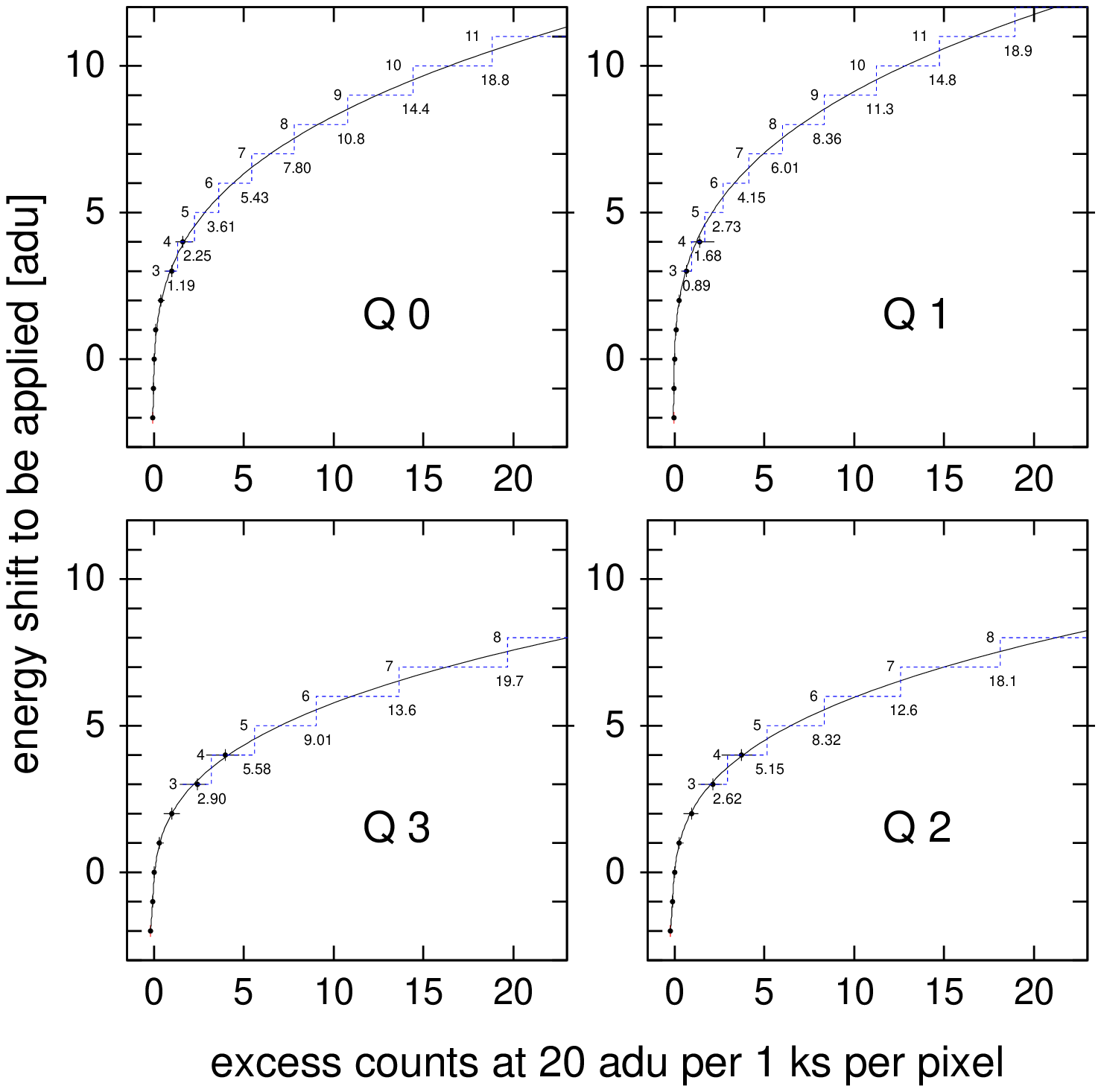,clip=,width=0.5\hsize,%
       bbllx=100pt,bblly=222pt,bburx=533pt,bbury=653pt}
\hfil
\raisebox{20mm}{\vbox{\hsize=0.4\hsize
\caption{Extrapolation of the empirical correlation between excess counts
in the 20~adu image and the energy shift to be applied. The numbers
below the horizontal steps specify the boundaries of the interval of excess
counts, where the energy scale should be corrected by the adu values given
to the left of the horizontal steps.
The extra\-polation was done by fitting a cubic function to the median values
of the excess counts found for offset shifts between $-2$ and $+4\mbox{ adu}$
(cf.\,Fig.\,\ref{offmp9_3_}); these points are marked by small dots.
An extrapolation was necessary because corrections by more than 4~adu
may occur, but they do not occur frequently enough for a direct determination
of the correlation function.
\label{offmp9_4}
}}\hfil}}
\end{figure}
%-------------------------------------------------------------------------------

\smallskip
However, not all of the offsets in the residual offset map are artefacts
caused by energetic particles. Comparison of such maps (e.g.\
Figs.\,\ref{offset_prep_526}c and \ref{offset_comp}a) indicates that they
consist of a superposition of both persistent and temporally variable
patterns. By computing for each pixel the median value of several such maps,
only the persistent patterns remain (Fig.\,\ref{offset_comp}b). Subtraction of
such a ``residual offset reference map'' from an individual residual offset
map reveals then the temporally variable content of this ``cleaned offset map".
Comparison of such maps with images accumulated from the low--energy events in
the subsequent exposure (see Fig.\,\ref{offset_comp}a and
Sect.\,\ref{sect:without_offset_map}) shows a clear correlation between the
position of the bright patches in the low--energy image and the pixels which
have a non--zero value in the cleaned offset map. This shows that the patches
seen in the cleaned offset maps reflect incorrect offsets caused by energetic
particles.

\smallskip
As the processed offset maps contain directly the adu shifts which have to be
subtracted from all events in the corresponding pixels in order to reconstruct
the correct energy scale, an implementation of this algorithm is
straightforward. However, while the correct energy scale can be reconstructed,
there is no possibility to recover events which were not transmitted to ground
because a too high offset was subtracted so that they fell below the lower
energy threshold. On the other hand, events where a too low offset was
subtracted will show up after the correction with adu values below the lower
threshold. In order to restore a homogeneous data set, these events should be
removed afterwards. Then the gain and CTI correction should be repeated for
the whole data set.

\medskip
\subsubsection{If the offset map is not available}
\label{sect:without_offset_map}

\smallskip
Even if the offset map is not available (this was the general situation in the
past), the correction described above can be achieved, by extracting the
necessary information directly from the event file. This is possible because
the amount of noise events is a monotonically and steeply increasing function
towards lower energies (Fig.\,\ref{offmp1a}). As this function is also fairly
stable in time, the intensity of a pixel in the lowermost energy channel
transmitted (20~adu) is expected to be correlated with the shift in the energy
scale: if the energy scale in a pixel is shifted by only 1~adu towards higher
energies, i.e., if the 20~adu bin contains 19~adu events, then this pixel
should contain $\sim1.8$ times more events.

\smallskip
However, this correlation is disturbed by the fact that the brightness of a
pixel at 20~adu is also influenced by other factors, in particular by its
individual noise properties. In Fig.\,\ref{ima20adu_comp}a, the lower half of
the detector (Q\,2 and Q\,3) are brighter than the upper one (Q\,0 and Q\,1).
Comparison with Fig.\,\ref{offmp1a} shows that this behaviour is caused by the
fact that there are more noise events at 20~adu in Q\,2 and Q\,3 than in Q\,0
and Q\,1. In addition to the brightness differences from quadrant to quadrant,
brightness variations can also be seen within the quadrants and even within the
CCDs (Fig.\,\ref{ima20adu_comp}a).

\smallskip
Separation of brightness variations caused by shifts in the energy scale from
variations caused by other effects can be achieved with a similar method as
applied to the residual offset maps (Sect.\,\ref{sect:with_offset_map}): from
a set of 20~adu images, derived from long exposures with no bright sources in
the field of view and normalized to the same exposure time, a reference image
can be computed, containing for each pixel the median value, i.e., its nominal,
temporally constant, 20~adu brightness (Fig.\,\ref{ima20adu_comp}b). This
reference image can then be subtracted from an individual 20~adu image, in
order to derive the brightness variations characteristic for that particular
exposure. The resulting ``cleaned 20~adu image" (Fig.\,\ref{ima20adu_comp}c)
shows a clear spatial correlation with the cleaned offset map of the same
exposure (Fig.\,\ref{offset_comp}c). In addition to the spatial correlation,
there is also a direct correlation between the brightness of the pixels in the
cleaned 20~adu images (normalized to the same exposure time) and the cleaned
offset maps, as expected. This correlation (Fig.\,\ref{offmp9_3_}) is
different between the four quadrants, due to their different noise properties.
No such differences were found among the CCDs within a quadrant.

\smallskip
The correlation shown in Fig.\,\ref{offmp9_3_} was derived from seven long
observations of the Lockman hole ($\sim636\mbox{ ks}$ in total). This region
contains only faint sources, which cannot disturb the offset maps or the
20~adu images. Due to the long exposure times, the statistical uncertainties
in the number of excess counts at 20~adu are minimized. The statistical
quality is good enough to deduce the energy shifts to be applied from the
number of excess counts for the range from $-3$ to $+4\mbox{ adu}$. As larger
shifts do not occur frequently enough for a direct determination of the
correlation function, they were determined by extrapolation
(Fig.\,\ref{offmp9_4}).

\smallskip
With the correlation derived above, it becomes possible to reconstruct the
offset shifts directly from the event file, at least in an approximate way:
while offset shifts can occur only at discrete adu steps, the correspondence
between the 20~adu brightness and the value of the offset shift is not always
unique (Fig.\,\ref{offmp9_3_}). In addition, the presence of Poissonian noise
in the 20~adu images, in particular for short exposures, limits the
sensitivity for locating the pixels where a correction of the energy scale
should be applied. The sensitivity of locating such pixels can be increased by
rebinning four consecutive rows of the 20~adu image into one row each, in
order to utilize the information about how the offset map was calculated
on--board. Nevertheless, this method is less accurate than the technique
described in the previous section, but it does improve the overall
data quality and can be applied if no offset map is available.

%-------------------------------------------------------------------------------
\begin{figure}
\hbox to \hsize{\hfil
\psfig{file=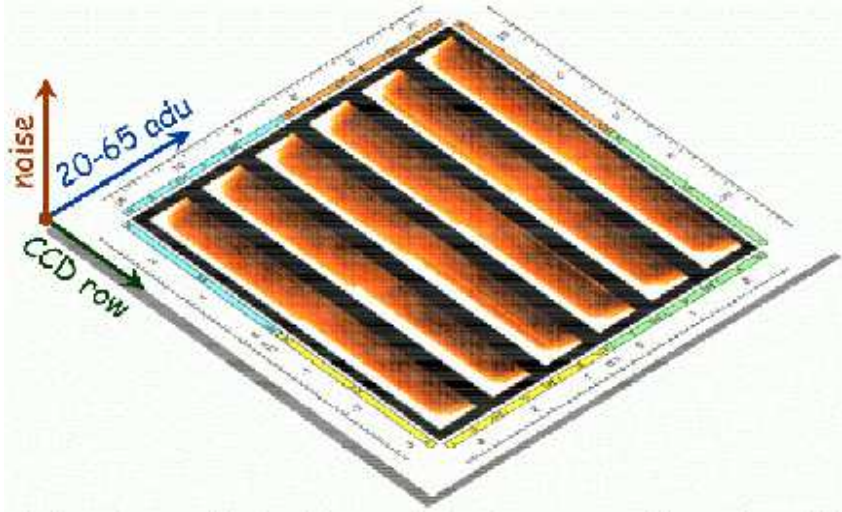,clip=,width=0.65\hsize}
\hfil
\raisebox{0mm}{\vbox{\hsize=0.3\hsize
\caption{Oblique view onto the EPIC pn detector, illustrating the spatial
variations of detector noise for the fullframe mode. For each row of each CCD,
the noise spectrum in the range 20\,--\,65~adu is displayed in a color code
which extends from 0 (black) to 0.01 events ks$^{-1}$ px$^{-1}$ adu$^{-1}$
(white); the maximum value is 1.3 events ks$^{-1}$ px$^{-1}$ adu$^{-1}$. In
order to cover the high dynamic range, the colors change with the square
root of the noise. These spectra were derived from 10 exposures with the filter
wheel closed (in XMM revolutions 129, 266, 339, 351, 363, 393, 409, 454, 462,
532), yielding a total exposure time of 144~ks. The information displayed here
is used for the suppression of the detector noise.
\label{noise_3d}
}}\hfil}}
\end{figure}
%-------------------------------------------------------------------------------

\bigskip
\section{DETECTOR NOISE}
\label{sect:detnoise}

\medskip
While there is practically no detector noise present at energies above
$80\mbox{ adu}$ ($\sim400\mbox{ eV}$), X-ray data below this energy, in
particular below 40~adu ($\sim200\mbox{ eV}$), are considerably contaminated
by noise events, which become more and more dominant towards the lowest
transmitted energy channels (Fig.\,\ref{offmp1a}). Investigations of 40~hours
of in--orbit calibration data with the filter wheel closed, taken over a
period of more than two years, show that the noise properties vary with
position and energy, but are fairly stable in time for most areas on the
detector. This property enables a statistical approach for suppressing the
detector noise: with the information about the spatial and spectral dependence
of the detector noise, the amount of events which correspond to the expected
noise can be flagged and suppressed.

\smallskip
The noise properties were derived from 10 exposures in fullframe mode, with the
filter wheel closed, between XMM--Newton revolutions \#129 and \#532, yielding
a total exposure time of 144~ks. These measurements were first corrected for
offset shifts (see Sect.\,\ref{sect:offset}) and then used for accumulating
raw spectra below 65~adu, individually for each CCD row
(Fig.\,\ref{noise_3d}). The fine spacing along readout direction was chosen
because the noise properties change considerably with distance from the
readout node, in particular close to the node. In order to get a sufficient
number of events, no subdivision was made along the CCD rows. This approach
was motivated by the fact that the noise properties do not show a pronounced
dependence along this direction (unless there is a bright column). As the
resulting spectra suffered somewhat from low count rate statistics, they were
smoothed along readout direction with a running median filter extending over
$\pm5$ rows. This smoothing was not applied to the 20 rows closest to the
readout nodes, where the spectra contained more counts and where the
dependence of the spectra on the distance from the readout nodes is high.

\smallskip
With this information, potential noise events in a specific observation can be
suppressed, on a statistical basis, in the following way: for each CCD row and
adu bin, the number of events is compared with the corresponding value in the
noise data (scaled to the same exposure). According to the ratio between the
actual number of events and the expected noise contribution, individual events
can then be randomly flagged. In order to improve the statistics somewhat, the
spectra from the observation to be corrected are internally smoothed by a
running median filter along readout direction before the noise contribution
is computed.

\smallskip
This method extends the usable energy range down to the lowest energies
transmitted, i.e.
$\sim120\,-\,140\mbox{ eV}$, depending on the position on the CCD. Below this
energy, parts of the detector become essentially insensitive due to the low
energy threshold applied on--board and the combined effect of charge transfer
loss and gain variations within the 768 amplifiers in the EPIC pn camera. In
addition to the improvement of the data quality, a removal of events which
were flagged as noise events also makes the files considerably smaller and
easier to handle.

\smallskip
Suppressing the detector noise by this method has not only advantages for
images, but leads also to a more reliable subtraction of detector noise in
the spectra and thus to an improved spectral quality. A large part of the
instrumental background can be taken into account by the unusal technique of
background subtraction, i.e., by determining the background spectrum from a
different region on the detector and subtracting it, appropriately scaled,
from the spectrum to be analysed. However, as the detector noise is not
vignetted (in contrast to background coming through the X--ray telescope) and
not distributed homogeneously across the detector (Fig.\,\ref{noise_3d}), the
usual background subtraction technique may lead to a systematic error in the
normalization of the low--energy part of the spectrum. Another interesting
consequence of the noise suppression method described here
for spectral studies is that the detector noise
can be separated from other background components.

%-------------------------------------------------------------------------------
\begin{figure}[b]
\vspace*{5mm}
\psfig{file=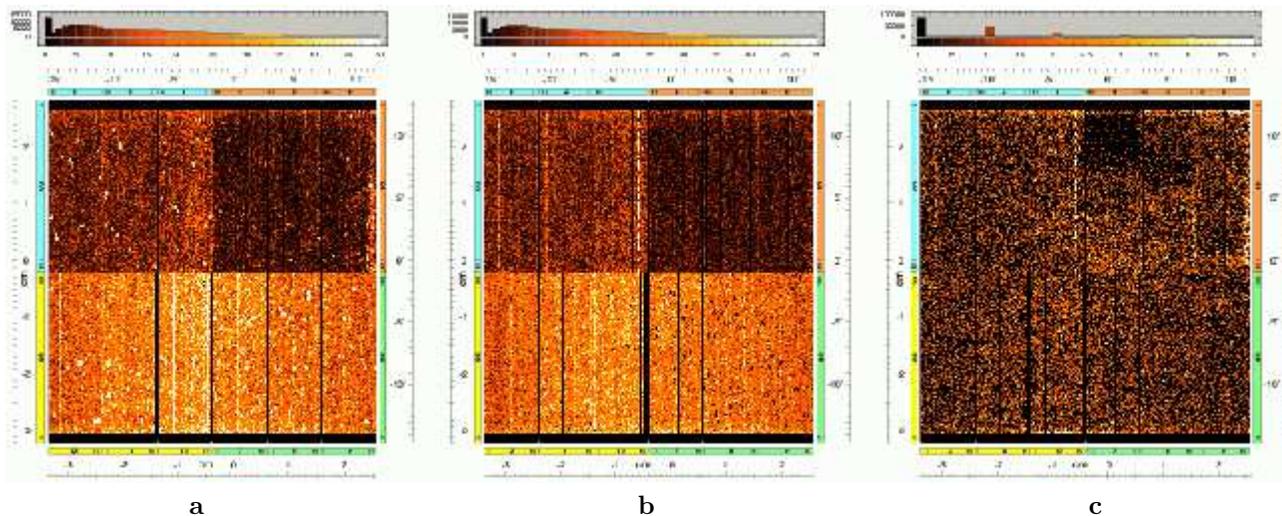,width=\hsize}
\hbox to \hsize{\bf \hfil a \hspace*{55mm} b \hspace*{55mm} c \hfil}
\caption{Illustration of the steps for correcting and cleaning the data set
for the 33.0~ks observation of the Vela SNR in XMM rev.\,367. Shown are images
accumulated from all events in the lowest energy channel transmitted (20~adu).
These images are dominated by detector noise (cf.\,Fig.\,\ref{offmp1a}),
despite the presence of the SNR. {\bf a)} Image derived from the original data
set. {\bf b)} Image after correcting the energy scale in specific pixels. {c)}
Image after suppressing also the detector noise. Note the different color
scale, which extends from 0 to 50 events per pixels for {\bf(a)} and {\bf(b)},
but from 0 to 5 events per pixels for {\bf(c)}.
\label{pn0367ff_med_0}
}
\end{figure}
%-------------------------------------------------------------------------------

%-------------------------------------------------------------------------------
\begin{figure}
\psfig{file=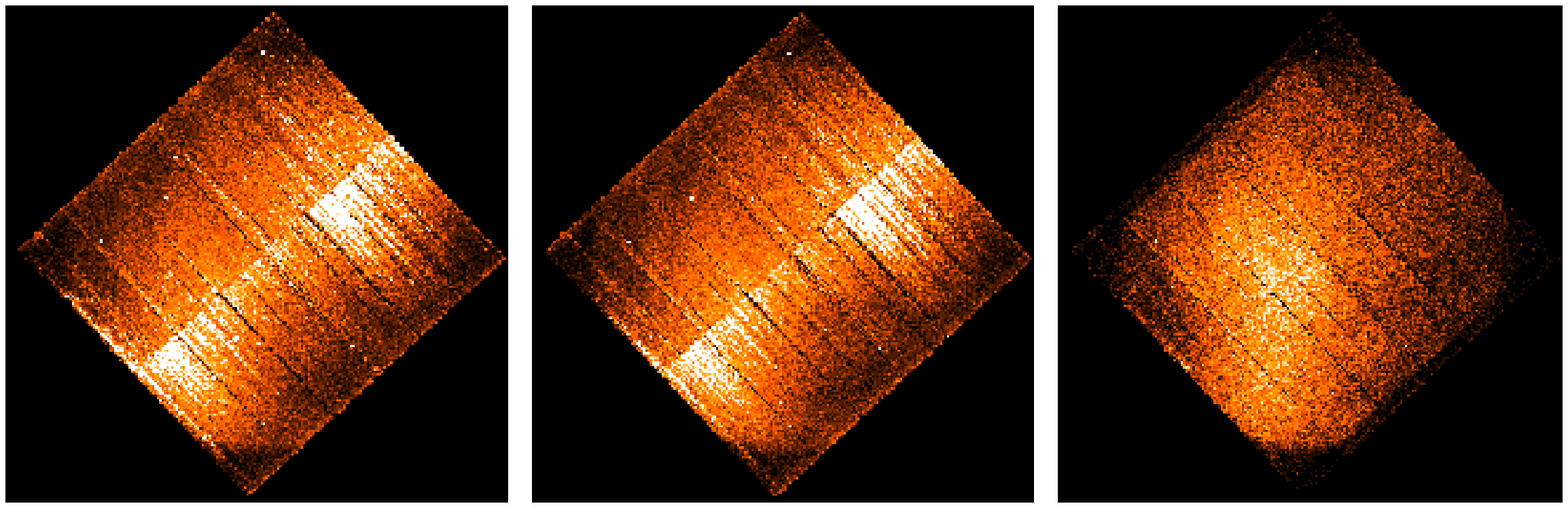,width=\hsize}
\hbox to \hsize{\bf \hfil a \hspace*{55mm} b \hspace*{55mm} c \hfil}
\caption{Effects of the steps illustrated in Fig.\,\ref{pn0367ff_med_0} on a
low--energy image of the Vela SNR. These image are all accumulated in
celestial coordinates (cf.\,Fig.\,\ref{vela_rosat_xmm}) in the (instrumental)
energy range 120\,--200~eV. {\bf a)} Original image, {\bf b)} after correcting
the energy scale in specific pixels, {\bf c)} after suppressing also the
detector noise. The color scale extends from 0 to 40 events per pixel for
{\bf(a)} and {\bf(b)}, and from 0 to 20 events per pixel for {\bf(c)}.
The original image {\bf(a)} contains $1.4\cdot10^6\mbox{ events}$, while the
cleaned image {\bf(c)} contains only $0.5\cdot10^6\mbox{ events}$.
\label{pn0367ff_med_1}
}
\vspace*{10mm}
\psfig{file=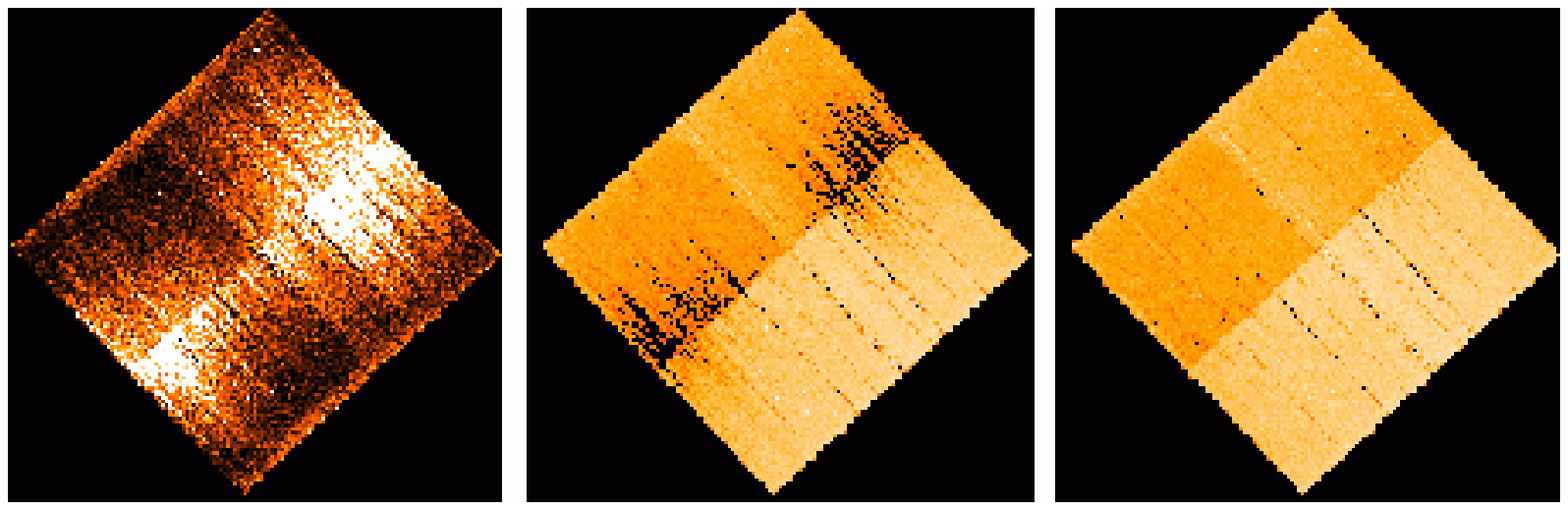,width=\hsize}
\hbox to \hsize{\bf \hfil a \hspace*{55mm} b \hspace*{55mm} c \hfil}
\caption{{\bf a)} Events in the 120\,--\,200~eV energy range which were
rejected as due to detector noise. The color scale extends from 0 to
20 events per pixel. This image contains $0.9\cdot10^6\mbox{ events}$
in total.
% 892\,713
{\bf b)} Image of all events with (instrumental) energies below 120~eV,
after correcting the energy scale in specific pixels, and displayed in
a logarithmic color scale extending from 0 to 500 events per pixel. This
image shows that not all areas of the detector are sensitive to energies
below 120~eV.
{\bf c)} Same as {\bf(b)}, but for energies below 140~eV. The insensitive
areas have practically all disappeared.
\label{pn0367ff_med_2}
}
\end{figure}
%-------------------------------------------------------------------------------

\medskip
\section{APPLICATION TO THE VELA SNR}
\label{sect:vela}

\begin{figure}
\hbox to \hsize{\hfil
\psfig{file=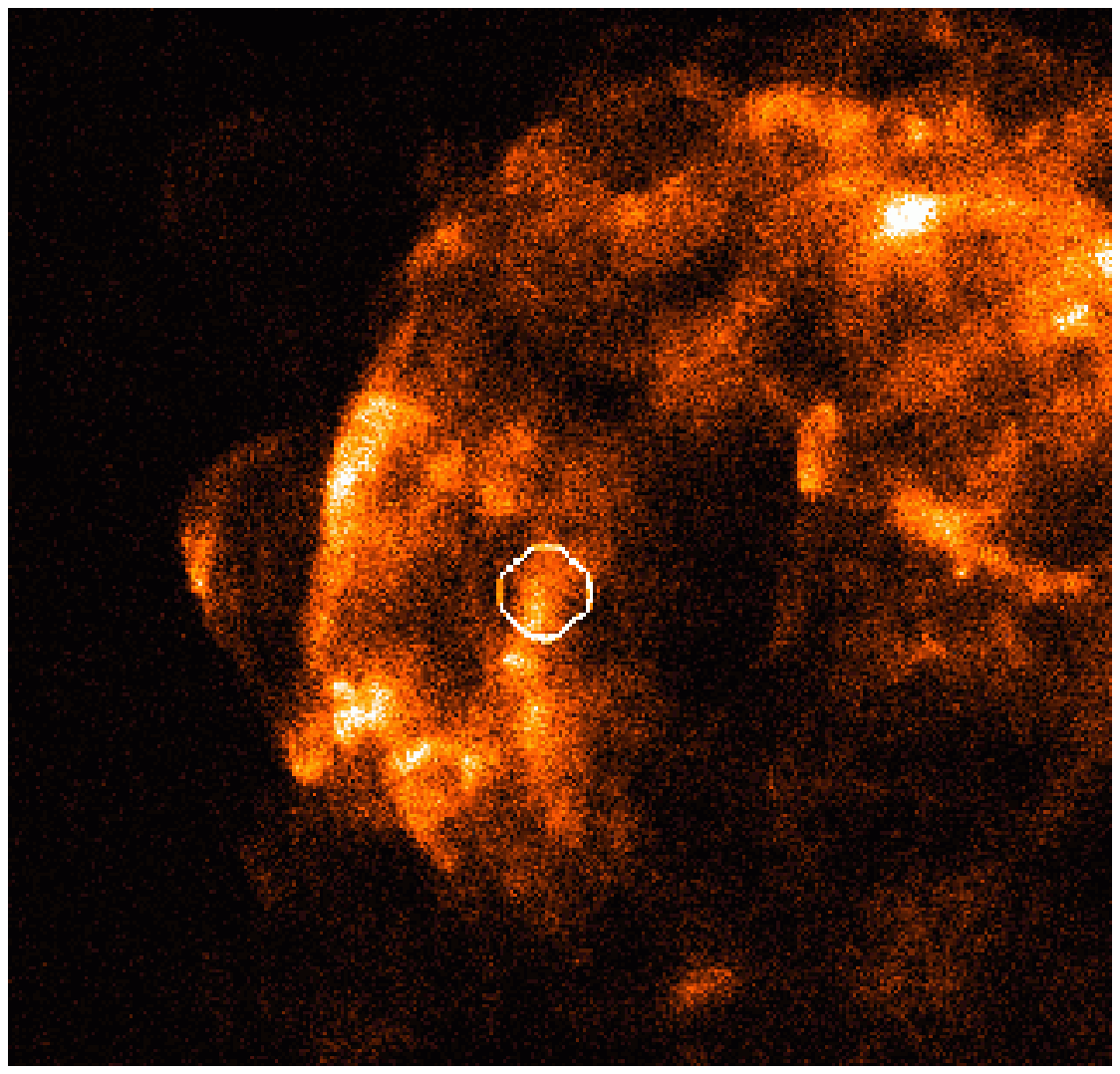,clip=,width=0.5\hsize}\hspace*{5mm}
\hspace*{-2mm}
\raisebox{20mm}{\vbox{\hsize=0.4\hsize
\caption{ROSAT 0.1\,--\,0.4~keV image of the eastern part of
the Vela SNR, obtained during the all--sky survey. The white contour
outlines the XMM--Newton EPIC pn field of view during the observation
in rev.\,367 (cf.\,Fig.\,\ref{vela_samples}).
\label{vela_rosat_xmm}
}}}}%
\vspace*{-6mm}%
\hbox to \hsize{\psfig{file=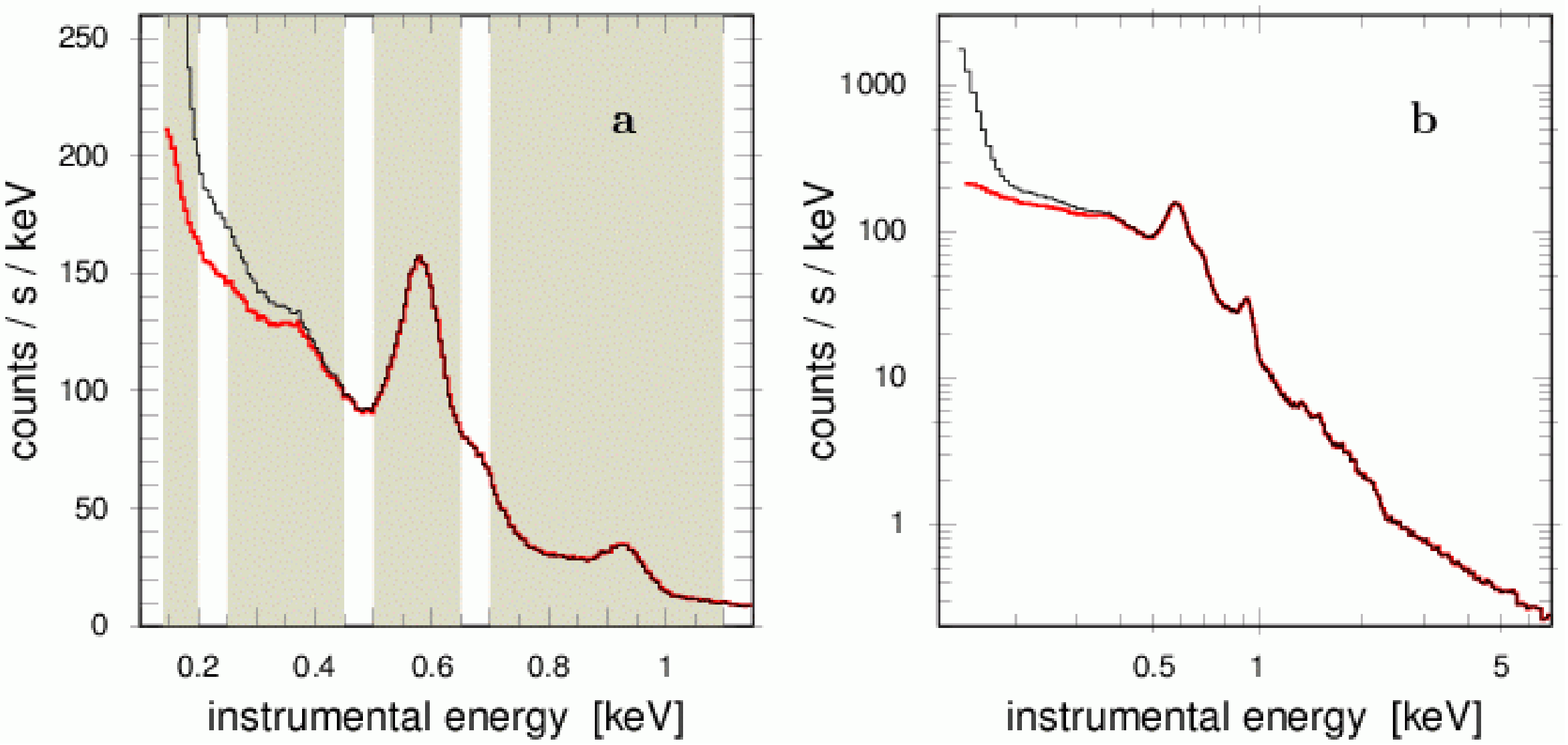,width=0.7\hsize}
\hspace*{-2.0cm}
\raisebox{0mm}{\vbox{\hsize=0.24\hsize
\caption{EPIC pn spectra of the full field of view from the
pointing to the Vela SNR, displayed in a linear {\bf(a)} and logarithmic
{\bf(b)} scale. The spectra are composed of all ``good'' ({\tt FLAG = 0}) and
valid patterns (singles, doubles, triples, and quadruples). The upper curves
show the original spectra, while the lower curves are the result of the
corrections described in the text. Shaded areas in {\bf(a)} mark the energy
bands which were selected for creating the images in Fig.\,\ref{vela_samples}.
\label{plspc7}
}}}}
\end{figure}

\medskip
Here we apply the methods described in Sections \ref{sect:offset} and
\ref{sect:detnoise} to a 33~ks
observation of the Vela supernova remnant, performed in December 2001 (XMM
rev.\,367) in fullframe mode with the medium filter. The observed region
(Fig.\,\ref{vela_rosat_xmm}) contains the north--west rim of another supernova
remnant, RXJ\,0852.0-4622, which was discovered in the ROSAT all--sky
survey\cite{98nat069}. Compared to the Vela SNR, which is
$\sim2\cdot10^4\mbox{ years}$ old, RXJ\,0852.0-4622 is considerably younger
(less than $1.5\cdot10^3\mbox{ years}$, with $\sim680\mbox{ years}$ as the
most probable value\cite{99aap296}) and is emitting more
energetic X--rays. In the ROSAT data, it becomes visible only above 1.3~keV.
At lower energies, it is outshone by the bright emission of the Vela SNR
(Fig.\,\ref{vela_rosat_xmm}).

\smallskip
Figure~\ref{pn0367ff_med_0}a shows the EPIC pn data of this observation
accumulated at 20~adu, the lowest energies transmitted. White patches
resulting from erroneous energy offsets are clearly seen. After applying the
local corrections of the energy scale described in Section~2, these patches
disappear (Fig.\,\ref{pn0367ff_med_0}b). The data, however, are still
dominated by detector noise. This noise can be suppressed by more than one
order of magnitude at 20~adu (Fig.\,\ref{pn0367ff_med_0}c) with the method
described in Section~3.

\smallskip
Although the 20~adu images are well suited for performing and verifying the
corrections, the spectral bandwidth is too small to see diffuse X--ray
emission from the Vela region. In order to demonstrate the improvement in the
low--energy X--ray data, Figs.\,\ref{pn0367ff_med_1}\,a-c show images for the
120\,--\,200~eV energy band, corresponding to the steps illustrated in
Figs.\,\ref{pn0367ff_med_0}\,a-c, The cleaned image
(Fig.\,\ref{pn0367ff_med_1}c) reveals that there is indeed diffuse X--ray
emission present, which is, however, lost in the detector noise if no
correction is applied (Fig.\,\ref{pn0367ff_med_1}a). The cleaning in this
energy range is substantial: 2/3 of the original events at 120\,--200~eV were
suppressed as due to detector noise. Fig.\,\ref{pn0367ff_med_2}a shows an
image of all the events which were been removed, in the same intensity scale as
Fig.\,\ref{pn0367ff_med_1}c.

\smallskip
While these methods make it possible to extend the useful energy range down to
$\sim120\mbox{ eV}$, there is a constraint which cannot be avoided: the
combined effects of charge transfer losses and variations in the gain of the
768 amplifiers cause an increase of the local low energy threshold (20~adu at
readout, corresponding to an instrumental energy of 100~eV) for specific areas
on the detector. This is illustrated in Fig.\,\ref{pn0367ff_med_2}b, which
demonstrates that not all areas of the detector are sensitive to energies below
120~eV. Homogeneous sensitivity across the whole detector is reached at
energies above $\sim140\mbox{ eV}$ (Fig.\,\ref{pn0367ff_med_2}c).

\begin{figure}
\psfig{file=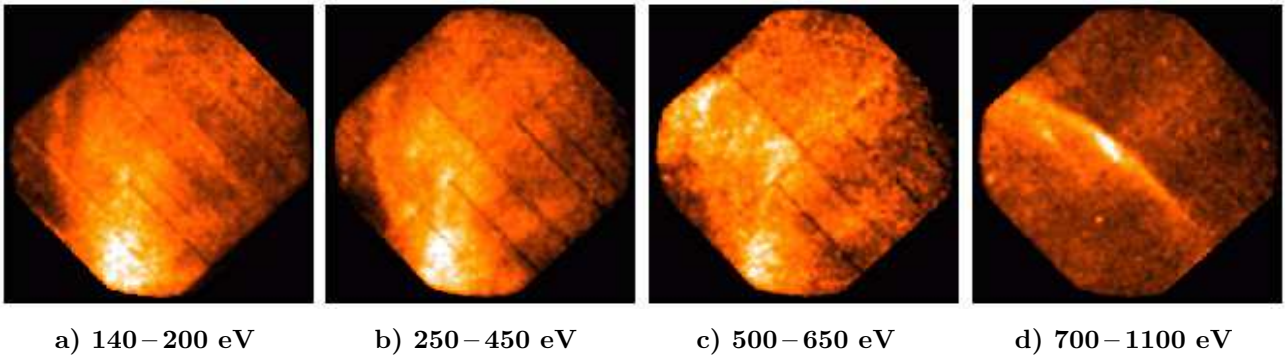,width=\hsize}
\vspace*{1mm}\hspace*{-7mm}\hbox to 18.7cm{\bf \hfil
a) 140\,--\,200 eV \hfil
b) 250\,--\,450 eV \hfil
c) 500\,--\,650 eV \hfil
d) 700\,--\,1100 eV \hfil}
\caption{EPIC pn images of the region of the Vela SNR containing the
north--west rim of RXJ\,0852.0-4622, observed in XMM rev.\,367
(cf.\,Fig.\,\ref{vela_rosat_xmm}). The energy ranges
indicated below each image are also shown in \,Fig.\,\ref{plspc7}a. All images
have been corrected for exposure variations and are composed of all valid
patterns, after having applied the processing steps described in the text.
Note the pronounced changes of the X--ray appearance in this small region of
the X--ray spectrum.
\label{vela_samples}
}
\end{figure}

\smallskip
All the images shown so far were obtained at energies below 200~eV. However,
the improvement in data quality is not restricted to this energy range.
Comparison of the spectra obtained before and after the correction
(Fig.\,\ref{plspc7}) indicates that the detector noise is suppressed up to
$\sim400\mbox{ eV}$ (corresponding to 80~adu), where it becomes negligible
(cf.\,Fig.\,\ref{offmp1a}). The high sensitivity of XMM EPIC pn makes it
possible to produce several narrow--band images of the Vela SNR region below
1~keV. Figure \ref{vela_samples} shows exposure corrected images derived from
the corrected and cleaned data set in the energy bands indicated in
Fig.\,\ref{plspc7}a.
These images reveal a stunning variety of complex X--ray emission, which
changes considerably with energy, even in this small 0.1\,--\,1.~keV spectral
range. Due to the improved spectral resolution of XMM compared to ROSAT,
emission from the north--west rim of the young supernova remnant
RXJ\,0852.0-4622, which is visible in the ROSAT data only above 1.3~keV, is
already obvious at energies above $\sim0.7\mbox{ keV}$.

%%%%% References %%%%%

%-------------------------------------------------------------------------------
%-------------------------------------------------------------------------------

%-------------------------------------------------------------------------------
%-------------------------------------------------------------------------------

\end{document}